\documentclass[10pt,A4]{article}
\usepackage{opex3_arXiv}
\usepackage{amssymb, amsmath}
\begin{document}
\title{Thin-disk laser scaling limit due to thermal-lens induced
  misalignment instability}

\author{Karsten Schuhmann$^{1,2,*}$,  
Klaus Kirch$^{1,2}$,
Francois Nez$^3$,\\
Randolf Pohl$^{4,5}$
and Aldo Antognini$^{1,2}$
}

\vspace{2mm}

\address{$^1$ Institute for Particle Physics, ETH, 8093 Zurich, Switzerland\\
  $^2$ Paul Scherrer Institute, 5232 Villigen-PSI, Switzerland\\
  $^3$ Laboratoire Kastler Brossel, UPMC-Sorbonne Universites, CNRS, ENS-PSL Research University,
   College de France, 4 place Jussieu, case 74 75005 Paris, France\\
  $^4$ Max Planck Institute of Quantum Optics, 85748 Garching, Germany\\
  $^5$ Johannes Gutenberg Universit\"at Mainz, QUANTUM, Institut f\"ur Physik\\
\& Exzellenzcluster PRISMA,  55099 Mainz, Germany
}

\vspace{2mm}
\email{$^*$ skarsten@phys.ethz.ch} 

\vspace{2mm}





\begin{abstract}
We present a fundemental obstacle in power scaling of thin-disk lasers related
with self-driven growth of misalignment due to thermal lens effects.
This self-driven growth arises from the changes of the optical phase
difference at the disk caused by the excursion of the laser eigen-mode
from the optical axis.
We found a criterion based on a simplified model of this
phenomenon which can be applied to design laser resonators insensitive
to this effect.
Moreover we propose several resonator architectures which are not
affected by this effect.
\end{abstract}




\section{Motivation}
\begin{figure}[t!]
\begin{center}
    \includegraphics[width=0.5\linewidth]{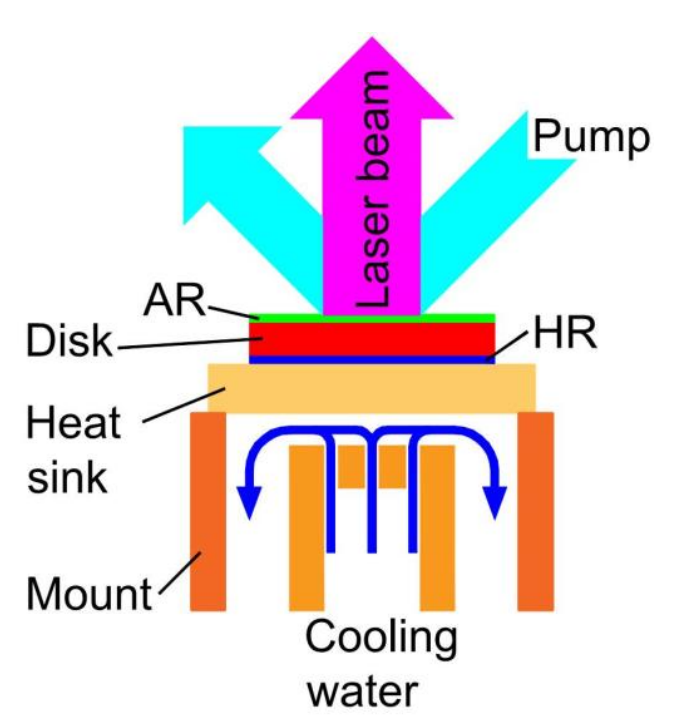}
\caption{\label{fig:TDL} (Color online) Scheme of the disk--heat-sink
  assembly (not to scale). The disk is thermally coupled to a
  water-cooled heat-sink. Cooling and temperature gradients occur
  along the disk axis which corresponds to the laser axis. The
  back and the front sides of the disk are coated with a high
  reflective (HR) and an anti-reflex (AR) layer, respectively for both
  pump and laser wavelengths.}
\end{center}
\end{figure}

Thin-disk lasers (TDL) are well know for their power scalability which
relates to the active medium geometry and its cooling
technique~\cite{Giesen1994, Brauch1995, Stewen2000}.
The laser crystal is shaped as a thin disk with a diameter of
typically several mm (depending on the output power/energy) and a
thickness of 100~$\mu$m to 400~$\mu$m, depending on the laser active
material, the doping concentration, the operation modus and the pump
design.
The backside of the disk is coated with dielectric layers
acting as high-reflector (HR) for the laser and the pump light, and it is contacted to a water-cooled heat
sink~\cite{Mende2009} as shown in Fig.~\ref{fig:TDL}.

As cooling occurs along the disk axis and pumping in quasi-end-pumped
configuration, the heat flow in the disk points along the disk axis
which is also the laser axis~\cite{Giesen1994, Giesen2007}.
The temperature gradients inside the laser crystal are thus mainly
parallel to the laser beam axis while in radial direction the
temperature within the homogeneously pumped central area is nearly
uniform.
Thermal lens effects are thus strongly suppressed in TDL because of
the efficient cooling (large surface  to volume ratio) and the small
temperature gradients in radial direction.
Another consequence of the disk--heat-sink geometry is that the
temperature  in the pumped region and therefore the thermal lens depend
only on the pump power density (assuming pump diameters larger than few
times the disk thickness).
This is one of the fundamental properties underlying
the power scalability of TDL.
However, power (energy) scaling calls for an increase of the beam
waist resulting in an increased sensitivity to the residual
thermal lens effects which eventually limits the achievable
scaling~\cite{Speiser2009, Speiser2007, Piehler2012, Mende2009a,
  Fattahi2014, Saraceno2015}.
When designing high power (energy) lasers therefore it is essential to
consider the stability properties of the resonator for variations of
the disk thermal lens~\cite{Speiser2007, Antognini2009, Zhu2014}.
These are usually represented in form of so called ``stability
plots''~\cite{magni1986} where the eigen-mode size at an optical
element in the resonator is plotted for variations of the disk
dioptric power $V$.

The thermal lens effect at the disk can be described using the
position-dependent optical phase difference OPD$(x,y)$ experienced by the laser
beam when reflecting on the disk.
The outgoing (after reflection) laser field amplitude $E_\mathrm{out}$
is given by
\begin{equation}
  E_\mathrm{out}(x,y)=  E_\mathrm{in}(x,y)\, e^{g(x,y)-i\frac{2\pi}{\lambda}\mathrm{OPD}(x,y)}
\end{equation}
where $E_\mathrm{in}$ is the in-going (before reflection) laser field
amplitude, $g(x,y)$ the space-resolved gain and $\lambda$ the laser wavelength.
A mathematical rigorous representation of the OPD can be accomplished
using Zernike polynomials~\cite{wang1980}.
Simplifying, here we represent the one-dimensional OPD as a Taylor
series
\begin{equation}
    \mathrm{OPD}(x)   =a + bx + cx^2+\cdots \;.
\end{equation}
Standard resonator designs implicitly assume an OPD of the form
OPD$\,=cx^2$, so that the disk can be described by a lens with focal
length of $f=1/(2c)$.
The linear term $bx$ is normally ignored because it simply describes
the tilt of a flat optical component which is accounted implicitly in
the alignment process of the laser resonator.
Similarly, the constant term $a$ produces a global phase shift which
corresponds to a change of the effective length of the resonator.
The dots represent higher-order contributions which have been widely
discussed in the literature~\cite{Piehler2012, Antognini2009,
  Siegrist1980} as they cause beam distortion and increased losses.

In this paper we consider in more detail the interplay between the
laser beam position at the active medium and the linear term $bx$.
As detailed later, a laser beam impinging on the disk with a given
deviation (excursion) from the disk--pumped-area axis induces a linear
term $bx$ in the OPD, i.e.,  a tilt of the disk.
This tilt causes a resonator response which further
modifies the position of the laser eigen-mode at the disk.
For certain resonator layouts a positive feedback between tilt and
laser beam position may exist which leads to a continuous growth of
the eigen-mode excursion at the disk resulting in a
disruption of the laser operation.
For other resonator configurations this interplay only leads to a
finite increase of the initial excursion implying a reduced
misalignment stability.
A parameter will be defined to easily identify  resonator designs
unstable with respect to this effect, whose importance increases with
the laser power.

\section{Thermal lens at the disk}

\begin{figure}[t!]
  \centering
    \includegraphics[width=1\linewidth]{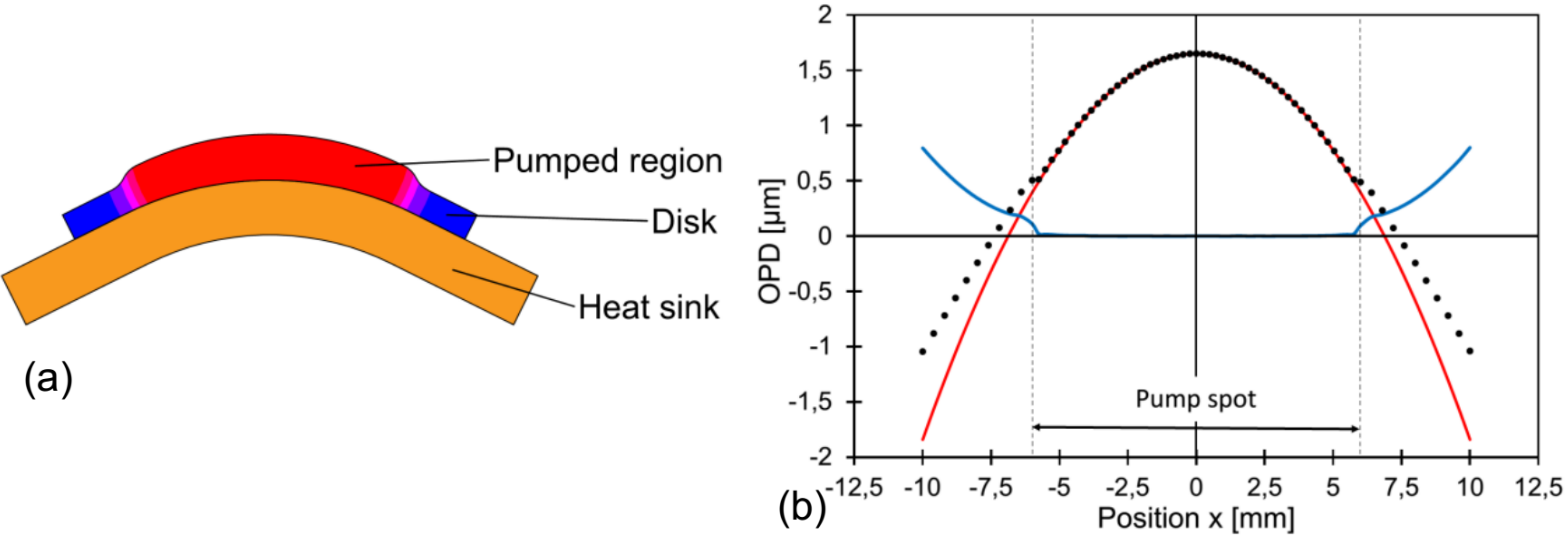}
\caption{\label{fig:pump} (Color online) (a) Scheme (not to scale) of
  the bending of the disk--heat-sink assembly due to the axial
  temperature gradient. The ``step-like'' profile in laser direction is
  induced by the thermal expansion related to the warmer temperature
  within the pumped region compared with the unpumped region. (b) The dots represent the
  optical phase difference (OPD) at the disk computed with FEM methods
  for the parameters as detailed in Appendix. Only the OPD caused by
  the pump process (fluorescence operation) is included here.  The red
  curve is a parabolic function fitted to the data in the central
  region $x\in[-3,3]$~mm. The residual between the fit and the 
  points is given by the solid blue line.}
\end{figure}

The rear side of the disk being at colder temperature expands
(in radial direction) less than the front side causing a bending of
the disk--heat-sink assembly as shown in Fig.~\ref{fig:pump}~(a).
We simulated OPD using finite element methods (FEM) to account for the
bending of the dielectric mirror (HR) at the backside of the disk
caused by this inhomogeneous radial expansion. These
simulations also account for the thermal expansion of the disk in the
laser direction and the variation of the disk refractive index versus
temperature $dn/dT$.

A typical OPD simulation which includes all these effects computed
using FEM as detailed in the Appendix is given in
Fig.~\ref{fig:pump}~(b).
Within the homogeneously pumped area the OPD can be well approximated
by a parabolic profile, while at the periphery of the pumped region
the OPD shows a deviation from the quadratic behavior which is
responsible for the excitation of higher-order beam components.

The OPD of Fig.~\ref{fig:pump}~(b) considers pump effects but neglects
thermal changes related to the laser operation.
Indeed laser operation reduces the heat deposition in the active
material as it increases the radiative de-excitation of the upper
laser levels at the expense of non-radiative
processes~\cite{Chenais2006, Chenais2004}.
Other mechanisms as a change of the effective quantum defect between
laser and fluorescence operation also contribute to this effect.
The model we will display in the following sections uses the effective
change of the thermal lens caused by the laser operation independently of
its origin.

\begin{figure*}[t!]
  \centering
  \includegraphics[width=1\textwidth]{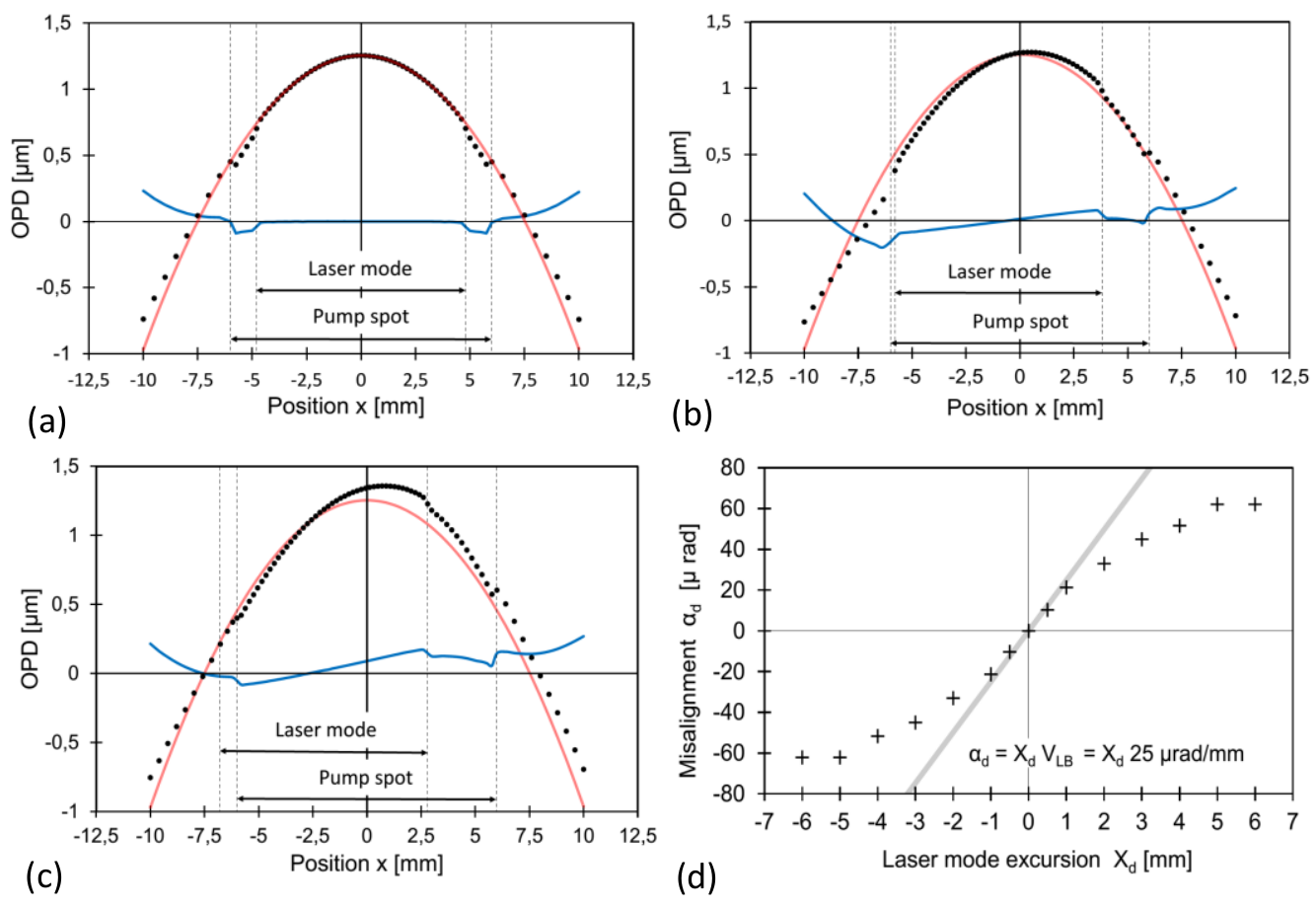}
\caption{\label{fig:OPD_pump_beam} (Color online) (a) The black points
  represent the one-dimensional optical phase difference (OPD) for a
  beam reflection at the disk computed with FEM methods as detailed in
  the Appendix. The pump and the laser beam share the same axis. The
  red curve is a parabolic function fitted to the FEM results in the central
  region $x\in[-3,3]$~mm. This curve is used as a reference in the
  following panels.  The difference between the fitted parabola and
  the simulated points is represented by the blue line. (b) Similar
  to (a) but in this case the laser beam impinges on the disk with an
  excursion of $X_d=1$~mm from the disk--pumped-area axis. The red curve is
  taken from (a).  (c) Similar to (b) with a beam excursion of
  $X_d=2$~mm. (d) The crosses represent the angular tilt $\alpha_d$ of the laser beam
  after a reflection at the disk caused by the change of the 
  thermal-lens due to the mode excursion $X_d$ computed with FEM methods.  The
  continuous line shows for comparison the prediction based on
  Eq.~(\ref{eq:tilt}) with $V_\mathrm{LB}$ obtained using the same FEM
  calculation. }
\end{figure*}

Figure~\ref{fig:OPD_pump_beam}~(a) shows the same FEM calculation as
displayed in Fig.~\ref{fig:pump}~(b) but now taking into account also
the reduction of the thermal load by a factor of
two~\cite{Chenais2006, Chenais2004, Perchermeier2013} due to laser
operation.
In this FEM computation the laser beam (resonator eigen-mode), the disk and
the pumped area have a common axis.
Within the laser eigen-mode  (which is smaller than the pumped area)
the OPD shows a quadratic behavior.
Hence, in this region the disk acts as a lens whose focal strength is
smaller than in Fig.~\ref{fig:pump}~(b) as expected from the decrease
of heat deposition in the laser mode.
The more complex structure at the periphery is related to the
superposition of the effect associated with the pump and the laser
mode which have been assumed to have different diameters.

The black points of Fig.~\ref{fig:OPD_pump_beam}~(b) show the OPD
computed with FEM when a laser beam (laser eigen-mode) impinges
on the disk 1~mm off-axis (in x-direction) relative to the
disk--pumped-area axis.
By subtracting from these points the quadratic function fitted to the
OPD where the laser beam and the pump spot are aligned (red curve in
panel (a)) a residual (blue continuous line) is obtained whose
central region shows a linear behavior.
This linear behavior can be interpreted as a tilt of the disk.
Therefore the FEM simulation indicates that a small excursion of the laser
mode from the disk--pumped-area axis induces in leading order only a
tilt of the disk, while the quadratic part (focal strength) remains unchanged.
This tilt grows with increasing laser eigen-mode excursion from the
disk--pumped-area axis as visible by comparing
Fig.~\ref{fig:OPD_pump_beam} (b) with (c) and as summarized in (d).

In the following we use a simplified model to show that 
an off-axis laser beam at the active medium induces a
change of the OPD that  can be well approximated by a
linear function with slope different from zero.
The FEM simulations have shown that the OPD caused by the pump beam
alone (fluorescence operation, see Fig.~\ref{fig:pump} (b)) and by the
pump beam combined with the lasing processes (see
Fig.~\ref{fig:OPD_pump_beam} (a)) in the vicinity of the disk axis can
be well approximated by parabolic profiles.
Thus we assume these OPDs to be of the form
\begin{eqnarray}
  \mathrm{OPD_{pump}}(x)\!\!\!&\!\!\!=\!\!\!&\!\!\!A_\mathrm{pump} +\frac{V_\mathrm{pump}}{2}x^2  \quad \mathrm{(fluorescence \;op.)}\\
  \mathrm{OPD_{tot}^{ONaxis}}(x)\!\!\!&\!\!\!=\!\!\!&\!\!\!A_\mathrm{tot} + \frac{V_\mathrm{tot}}{2}x^2 \quad  \mathrm{(laser \;operation)},
\end{eqnarray}
where $x$ is the variable describing the position relative
to the optical axis, $V_\mathrm{pump}$ and $V_\mathrm{total}$ are the
focal strengths of the parabolic profiles, and $A_\mathrm{pump}$,
$A_\mathrm{tot}$ constants describing a global (position-independent)
phase shift.
The difference between these two OPDs is used to define the OPD
arising from the laser beam (LB) only
\begin{eqnarray}
  \mathrm{OPD_\mathrm{LB}}(x) &=& \mathrm{OPD_{tot}^{ONaxis}}(x)-\mathrm{OPD_{pump}}(x)\\
  &=& (A_\mathrm{tot}- A_\mathrm{pump}) + \frac{V_\mathrm{tot}-V_\mathrm{pump}}{2}x^2.
\end{eqnarray}
For TDL the thermal lens is usually defocusing because it is dominated
by the bending of the disk as shown in Fig.~\ref{fig:pump} (a) so that the
disk acts as a convex mirror.
Consequently, for TDL the  focal strength associated with the laser beam only 
$V_\mathrm{LB}=V_\mathrm{tot}-V_\mathrm{pump}$ is positive (focusing)~\footnote{
For rod lasers the thermal lens is usually focusing because
the refractive index change versus temperature is positive
($dn/dT>0$).
Hence, the reduced heat load due to laser operation leads to a negative
(defocusing) $V_{LB}$.}.

As a next step we consider the OPD resulting from the pump process and
a laser beam impinging off-axis on the active medium:
\begin{eqnarray}
  \mathrm{OPD_{tot}^{OFFaxis}}(x) &=& \mathrm{OPD_{pump}}(x)+\mathrm{OPD_\mathrm{LB}}(x-X_d)\\
  &=& A_\mathrm{tot} + \frac{V_\mathrm{pump}}{2}x^2+\frac{V_\mathrm{LB}}{2}(x-X_d)^2
\end{eqnarray}
where $X_d$ is the excursion of the laser beam from the optical axis.
A misalignment of the laser beam by $X_d$ from the disk--pumped-area
axis thus gives rise to an OPD variation given by
\begin{eqnarray}
  \Delta(\mathrm{OPD})(x) &=&\mathrm{OPD_{tot}^{ONaxis}(x)-OPD_{tot}^{OFFaxis}}(x) \\
  &=& +\frac{V_\mathrm{LB}}{2}X_d\,x - \frac{V_\mathrm{LB}}{2}X_d^2.
  \label{eq:delta-OPD}
\end{eqnarray}
The last term in Eq.~(\ref{eq:delta-OPD}) is a position-independent
contribution which describes an overall phase shift that can be neglected in
our treatment.
The first term being linear in $x$ represents the angular tilt $\alpha_d$ suffered
by a laser beam after reflection at the disk
\begin{equation}
  \alpha_d=+V_\mathrm{LB}X_d.
  \label{eq:tilt}
\end{equation}
Therefore, the simplified model predicts that a beam excursion $X_d$
at the active medium induces a tilt of $\alpha_d/2$ of the active medium
proportional to the beam excursion.
As well visible in Fig.~\ref{fig:OPD_pump_beam}~(d) for small
excursions ($X_d\lesssim 1$~mm) there is a good agreement between the tilt
calculated using only the FEM and the tilt based on
Eq.~(\ref{eq:tilt}) with $V_\mathrm{LB}$ also from the same FEM.
For larger excursion this agreement decreases.

Equation~(\ref{eq:tilt}) represents the steady state tilt of the disk
caused by a fixed beam excursion.
It also implies that a change of the beam excursion causes
a change of the disk tilt.
However, the adjustment of the disk tilt to the new beam excursion
is not instantaneous but occurs with a time constant $\tau$
given by the thermalization of the disk--heat-sink assembly.
For example the variation of the temperature distribution for the
disk--heat-sink assembly presented in the Appendix has been computed
to have a time constant of $\tau=5$~ms.
Due to the linearity of Eq.~(\ref{eq:tilt}) we can model the time
variation of the disk tilt as\footnote{Here we implicitly assumed that
  there is only a single thermalization time $\tau$, i.e., that $\tau$
  does not depend on the ($x$, $y$) position because the heat flow in
  the active material occurs in axial ($z$) direction. Diamond as
  substrate material shows also a radial heat flow given its large
  thermal conductivity (see Table~\ref{tab3}). However, this large
  thermal conductivity and the low thermal expansion leads to a
  thermal lens effects order of magnitude smaller than the one
  generated by the active material. Therefore, to a good
  approximation, only the heat flow in the laser crystal has to be
  considered which is along the $z$ axis due to the moderate thermal
  conductivity of the active material and its small thickness. For
  diamond substrates the above assumption is thus justified.  A
  metallic heat sink has a significant lower thermal conductivity
  compared to diamond leading to a negligible heat flow in $x$ and
  $y$ directions (assuming large pump spots). Hence, the assumption of
  a single thermalization time $\tau$ is fulfilled also for metallic
  heat sinks.}
\begin{equation}
  \tau \frac{d\alpha_d(t)}{dt}=V_\mathrm{LB}X_d(t)-\alpha_d(t).
    \label{eq:tilt_vs_time2}
\end{equation}

\section{Resonator reaction for end-mirror misalignment}

A geometrical ray propagating in an optical system 
can be described by its position $X_r(z)$ and its angle $\theta_r(z)$ with
respect to the optical axis (z-axis)~\cite{Siegman1986}.
For an ideally aligned optical system the beam propagates along
the optical axis of the system  so that $X_r(z)=0$ and
$\theta_r(z)=0$ are fulfilled everywhere.
The ABCD-matrix formalism can be used to compute the beam propagation
along the optical system if the initial beam position and angle are known.
In a resonator the eigen-mode  has to reproduce itself after a round-trip
with regard to its position, angle, waist and  phase front curvature.

Starting from an ideally aligned laser resonator where the laser
eigen-mode is on-axis everywhere, we introduce a small misalignment of
the first end-mirror by an angle $\alpha_r/2$.
To have laser operation the eigen-mode position $X_r$ and the angle
$\theta_r$ at the first end-mirror must fulfill following equation
\begin{equation}
  \left[
  \begin{array}{c}
     X_r \\
     \theta_r
  \end{array} \right]= 
  \left[
  \begin{array}{ c c }
     D & B \\
     C & A
  \end{array} \right]
  \left[
  \begin{array}{ c c }
     A & B \\
     C & D
  \end{array} \right]
  \left[
  \begin{array}{c}
     X_r \\
     \theta_r+\alpha_r
  \end{array} \right] \label{eq:matrix}
\end{equation}
where the second ABCD-matrix describes the beam propagation from the
first (tilted) end-mirror to the second end-mirror, and the first
ABCD-matrix the back-propagation from the second end-mirror to the
first end-mirror.  Note that the effective focal strength (thermal and
non-thermal) of the disk is included in the two matrices.
Using the condition that  the determinant of each ABCD-matrix is
equal to one, the solution of these equations reads
\begin{eqnarray}
  X_r &=& -\frac{\alpha_r}{2} \frac{D}{C}  \label{eq:X_r}\\
  \theta_r &=& -\alpha_r/2. \label{eq:theta_r}
\end{eqnarray}

Because the resonator reaction time (10~ns time scale) is much shorter than the
thermal lens adaptation time (ms time scale) we can assume that
\begin{equation}
  X_r(t)=-\frac{\alpha_r(t)}{2} \frac{D}{C}
  \label{eq:tilt_vs_time}
\end{equation}
holds for any time $t$.

Equation~(\ref{eq:tilt_vs_time}) describes the time-dependent beam
excursion at the resonator end-mirror that results as a
consequence of the resonator response to a misalignment of the same
end-mirror by an angle $\alpha_r/2$.
This equation will be used in Sec.~\ref{sec:interplay} to model the
stability of a resonator having the disk as an end-mirror.
For a disk used as bending mirror (for V-shaped resonators) an
equation analogous to Eq.~(\ref{eq:tilt_vs_time}) has to be derived.
This is accomplished in Sec.~\ref{sec:interplay2} and applied to
resonator stability studies in Sec.~\ref{sec:impact}.

\section{Resonator stability for disk as end-mirror}
\label{sec:interplay}

We have seen previously that an excursion of the eigen-mode at
the disk position causes a change of the disk tilt (thermal lens
effect), and that a tilt of the disk causes an excursion of the
eigen-mode at the disk position (resonator reaction).
Till now we neglected the interplay of these two effects.
Their coupling causes a feedback loop which calls
for a more detailed investigation.

Coupling of these two effects is realized by identifying
\begin{eqnarray}
  X & \equiv & X_d = X_r+X_0\label{eq:X}\\
  \alpha & \equiv& \alpha_d = \alpha_r \label{eq:alpha}.
\end{eqnarray}
Here we have assumed an initial eigen-mode excursion $X_0$ at
the disk position.
The magnitude of this excursion is unimportant for the understanding
of the effect we are modeling in this study as will become clear
later (see Eq.~(\ref{eq:time-behaviour})).
It must simply have a non-vanishing value: $X_0\neq 0$.
Note that this condition is always valid in practice because of the
imperfection (misalignment between pump and laser mode) intrinsic in
the alignment of the resonator.

\begin{figure}[t!]
  \centering
  \includegraphics[width=0.7\linewidth]{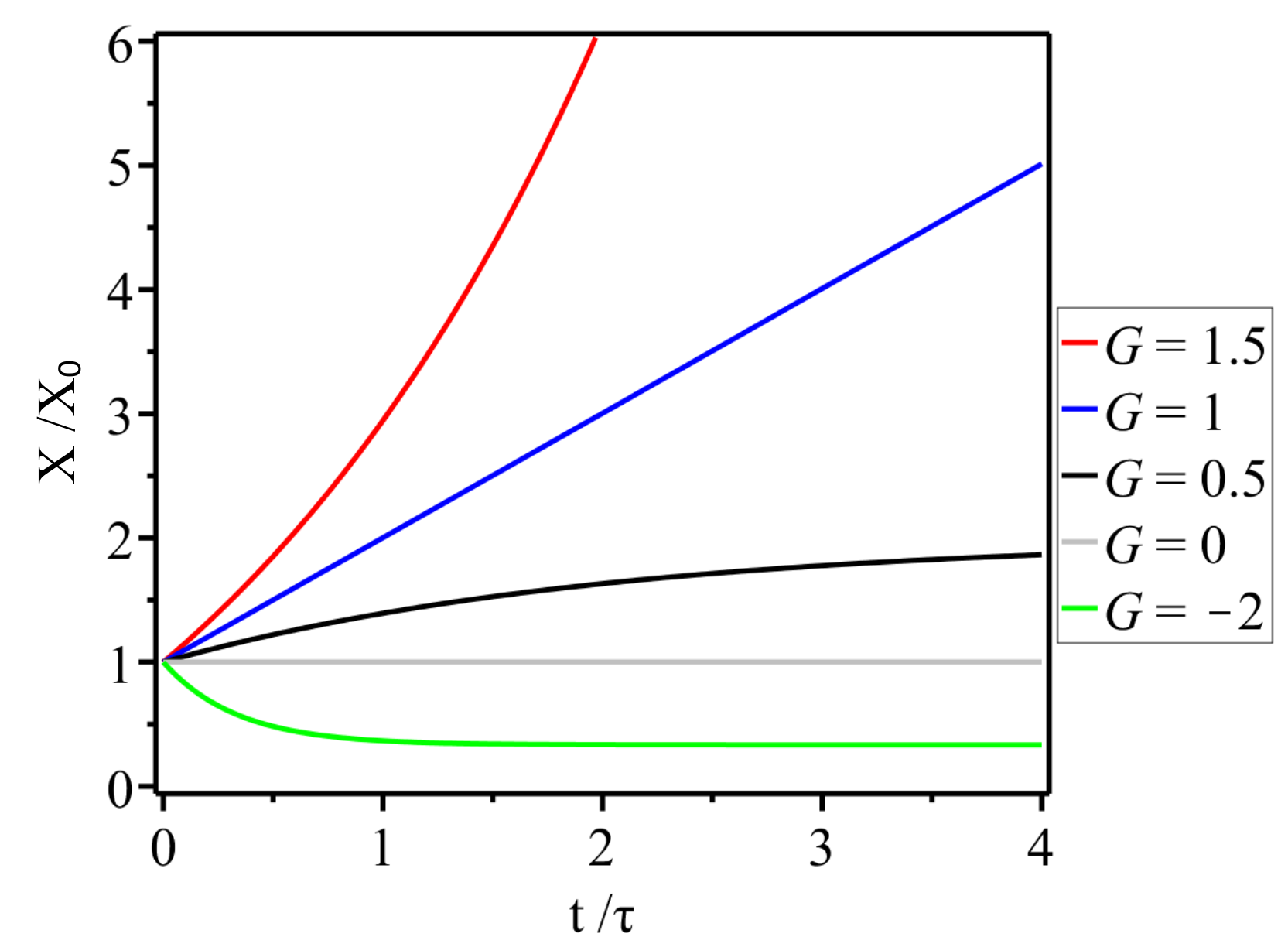}
\caption{\label{fig:time-evolution} (Color online) Time evolution of
  the eigen-mode excursion  from the
  disk--pumped-area axis caused by the interplay between resonator
  reaction and change of the thermal lens due to misalignments. An initial
  excursion $X_0$ can be either reduced or amplified to a
  finite or infinite value depending on the value of the $G$ parameter.  }
\end{figure}

By combining Eqs.~(\ref{eq:tilt_vs_time2}), (\ref{eq:tilt_vs_time}),
(\ref{eq:X}) and (\ref{eq:alpha}) and assuming $dX_0/dt=0$ (static or
slowly varying initial misalignment) we obtain
\begin{equation}
  \tau \frac{dX(t)}{dt}=GX(t)-[X(t)-X_0]
  \label{eq:differential}
\end{equation}
where we have defined the parameter $G$ as
\begin{equation}
  G=-\frac{V_\mathrm{LB}}{2}\frac{D}{C}.
  \label{eq:G}
\end{equation}
The solution of Eq.~(\ref{eq:differential}) reads:
\begin{equation}
  X(t)=\left\{
  \begin{array}{ l l }
     X_0\frac{ G\cdot\exp{\big(\frac{G-1}{\tau}t}\big)-1}{G-1} & \mbox{for $G \neq 1$}  \\[1mm]
     X_0\big(1+\frac{t}{\tau}\big) & \mbox{for G=1}.
  \end{array}\label{eq:time-behaviour}
  \right.
\end{equation}  
The time behavior $X(t)$ for various values of $G$ is plotted in
Fig.~\ref{fig:time-evolution}.
For $G<0$ the initial excursion is reduced with time, for $G=0$ it
remains constant, and for $G>0$ it is amplified.
Furthermore, for $G<1$ the initial excursion is amplified but
saturates with time to a finite value, while for $G \geq 1$ the
initial excursion increases continuously, $X(t\rightarrow
\infty)=\infty$.

The rate of change of the excursion $\frac{dX(t)}{dt}$ depends on $G$,
$X_0$ and $\tau$.
Yet the fate of the  excursion at large times $X(t\rightarrow \infty)$,
i.e., whether it remains constant, damped or amplified depends only on
the parameter $G$.

We close this section by listing the various assumptions and 
limits underlying the analytical solution of the time evolution of the
eigen-mode excursion given in Eq.~(\ref{eq:time-behaviour}).
A linear dependence between excursion and tilt has been assumed which
is valid only for small excursions ($X\lesssim 1$~mm) as demonstrated in
Fig.~\ref{fig:OPD_pump_beam}~(d).
We also assume a space independent time constant $\tau$.
Deviation from this behavior impact in a minor way our model as $\tau$
does not affect the fate of the beam excursion.

In this model we neglect  the
decrease of the circulating laser intensity caused by a misalignment.
On one hand, this intensity decrease reduces $V_\mathrm{LB}$ and the tilt of
the disk, on the other hand, the ratio $D/C$ which depends on the
thermal lens increases making this resonator more unstable.

In principle our model could be extended to include all these effects
and the soft-aperture effects naturally occurring in the pumped
medium~\cite{Siegman1986, Schuhmann2016}.
This would lead to a very complex interplay obscuring the principle of
the mechanism we are disclosing in this study and whose precise
modeling goes beyond the scope of this paper.
However, our simplified analytical  model captures correctly the onset of
this misalignment instability that causes a dramatic decrease of the  laser
performance (efficiency and stability).
Thus its predictive power for designing resonator remains unaffected.

\section{Stability of V-shaped resonators}
\label{sec:interplay2}

In this section we investigate the stability properties of V-shaped
resonators (widely used in the TDL sector~\cite{Brauch1995, Antognini2009, Brunner2002, Paschotta2001, Baer2010}) with respect to
the thermal-induced misalignment effect disclosed in this paper.
More specifically we model here the eigen-mode excursion at the disk for a
resonator where the disk is a folding mirror (not an end-mirror).

The round-trip propagation in this resonator can be divided into two
branches (right and left of the disk): from the tilted disk to the
second end-mirror and back to the disk
\begin{equation}
  \left[
  \begin{array}{c}
     X'_r \\
     \theta'_r
  \end{array} \right]= 
  \left[
  \begin{array}{ c c }
     D_R & B_R \\
     C_R & A_R
  \end{array} \right]
  \left[
  \begin{array}{ c c }
     A_R & B_R \\
     C_R & D_R
  \end{array} \right]
  \left[
  \begin{array}{c}
     X_r \\
     \theta_r+\alpha_r
  \end{array} \right], \label{eq:matrix1}
\end{equation}
and from the tilted disk to the first end-mirror and back to the disk
\begin{equation}
  \left[
  \begin{array}{c}
     X_r \\
     \theta_r
  \end{array} \right]= 
  \left[
  \begin{array}{ c c }
     A_L & B_L \\
     C_L & D_L
  \end{array} \right]
  \left[
  \begin{array}{ c c }
     D_L & B_L \\
     C_L & A_L
  \end{array} \right]
  \left[
  \begin{array}{c}
     X'_r \\
     \theta'_r+\alpha_r
  \end{array} \right]. \label{eq:matrix2}
\end{equation}
%
%
The ABCD-matrixes describe the left ($L$) and the right ($R$) branches of
the optical system similar to Eq.~(\ref{eq:matrix}).
The effective focal strength of the disk is included in the ABCD
matrixes: it can be included without loss of generality either in the
left or in the right branches or even divided between the two branches.
$X_r$ and $\theta_r+\alpha$ represent the excursion and the angle for
the beam leaving the tilted disk  towards the second (right) end-mirror.
We assume the disk to be tilted by an angle $\alpha_r/2$.
$X'_r$ and $\theta'$ are the excursion and angle of the beam returning
to the disk (prior to reflection on the disk) after reflection on
the second end-mirror, i.e., after a propagation in the right branch.
The beam leaving the disk toward the first (left) end-mirror after a
reflection on the disk has thus an excursion and angle of $X'_r$ and
$\theta'+\alpha_r$, respectively.
When it returns back at the disk after a propagation in the left branch
it has an excursion $X_r$ and an angle $\theta_r$.

The eigen-mode excursion at the disk position caused by a tilt of the
disk by an angle $\alpha_r/2$ can be found by solving these two
coupled equations and reads
\begin{equation}
  X_r = -\alpha_r \frac{A_LD_R}{A_LC_R+C_LD_R}  .   \label{eq:X_r_2}\\
\end{equation}
Following a similar argumentation as exposed in
Sec.~\ref{sec:interplay} we find that the time evolution of the
excursion $X(t)$ at the disk follows Eq.~(\ref{eq:time-behaviour}) but
with the parameter $G$ defined as
\begin{equation}
  G=-V_\mathrm{LB}\frac{A_LD_R}{A_LC_R+C_LD_R} .
  \label{eq:G2}
\end{equation}
On that account, all the conclusions drawn in previous section remain
valid after the appropriate replacement of the parameter $G$.

\section{Impact on typical V-shaped resonators }
\label{sec:impact}
\begin{figure}[t!]
  \centering
  \includegraphics[width=0.60\linewidth]{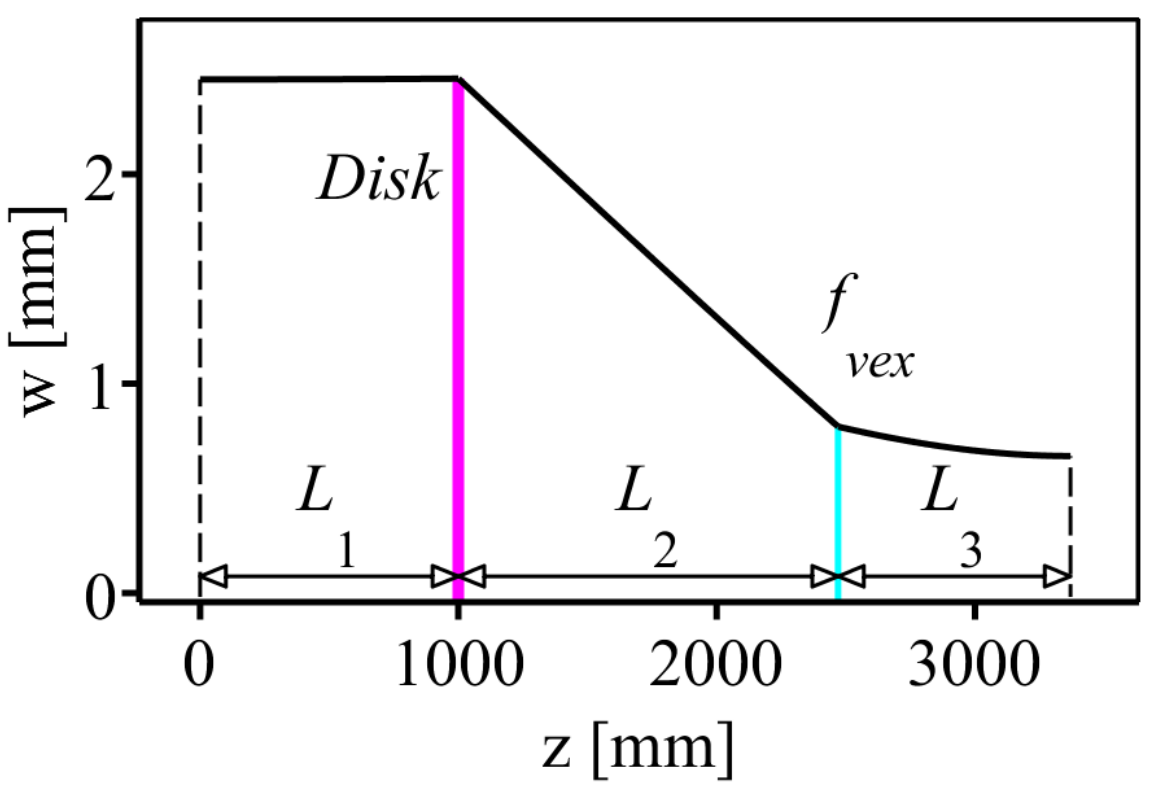}
\caption{\label{fig:layouts} (Color online) V-shaped resonator
  architecture consisting of a flat end-mirror, a free propagation of
  length $L_1$, a disk with 0.5~dioptric power (in unpumped
  conditions), a free propagation of length $L_2$, a convex lens
  (mirror) with focal length $f_\mathrm{vex}$, a free propagation of
  length $L_3$, and a flat end-mirror.  The eigen-mode waist w along the
  resonator for a particular set of values is also given to clarify
  the layout. }
\end{figure}

One of the first steps when designing laser resonators is to study the
influence of the thermal lens on the resonator stability.
This is achieved usually by means of so called ``stability
plots''~\cite{magni1986}, i.e., by plotting the evolution of the
eigen-mode size at a given optical element for variations of the
thermal lens.
Given a resonator layout, stable laser operation is achieved only in a
limited range of thermal lens values.
This range is known as the ``stability region''.

In this section we illustrate the shrinkage of the effective stability
region and the reduction of the output power caused by the
thermal-induced misalignment mechanism.
The impact of this misalignment mechanism will be illustrated for four
resonator layouts, all based on the architecture sketched in
Fig.~\ref{fig:layouts}.

\begin{figure}[t!]
  \centering
  \includegraphics[width=0.85\linewidth]{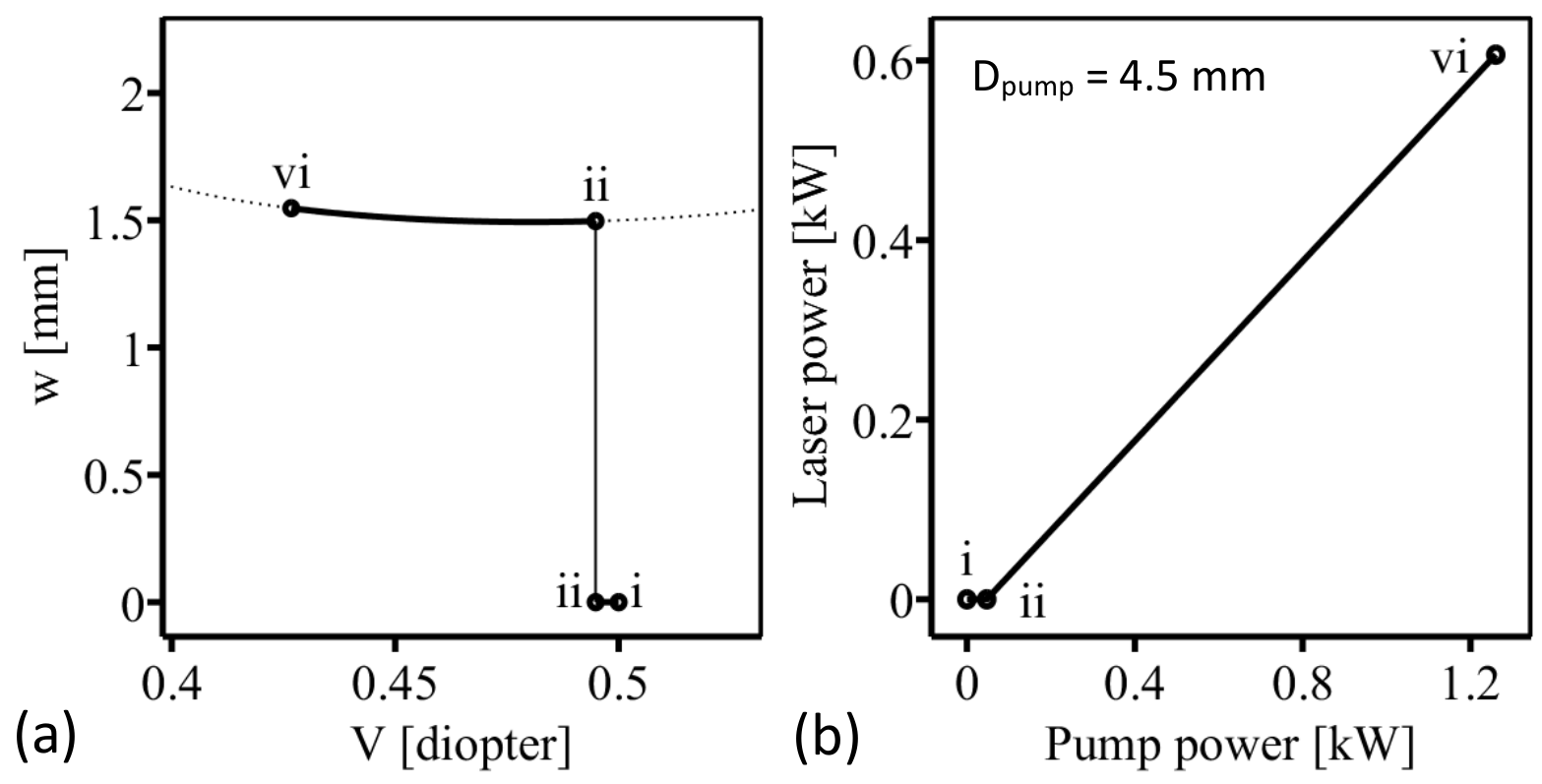}
\caption{\label{fig:stability_layout1} (a) Eigen-mode
  waist at the disk versus  the disk dioptric power $V$ for a
  resonator layout as in Fig.~\ref{fig:layouts} with $L_1=500$~mm,
  $L_2=1470$~mm, $L_3=100$~mm and $f_\mathrm{vex}=-1500$~mm. The
  dotted line represents the ``classical'' stability plot, i.e., the
  waist w for any  $V$.  The continuous black
  line indicates the beam waist versus dioptric power only for the
  dioptric power that pertain the disk in the assumed running
  conditions: starting from $V=0.5$~diopters for no pumping to
  $V\approx0.42$~diopters for 8~kW/cm$^2$.  The value of $\mathrm{w}=0$ is used to indicate
  that at the given dioptric power there is no laser operation.  The
  Roman numbers indicate specific states of the laser operation as
  described in the main text and summarized in Table~\ref{tab2}.
  Between (i) and (ii) there is no laser operation because the gain at the
  disk in this pump power density range is still smaller than the
  out-coupler transmission. (b) Qualitative evolution of the output
  power for the resonator layout of (a) as a function of the pump
  power.  The parameters of Table~\ref{tab1} have been used to
  model the output power evolution.}
\end{figure}

\begin{figure}[t!]
  \centering \includegraphics[width=0.85\linewidth]{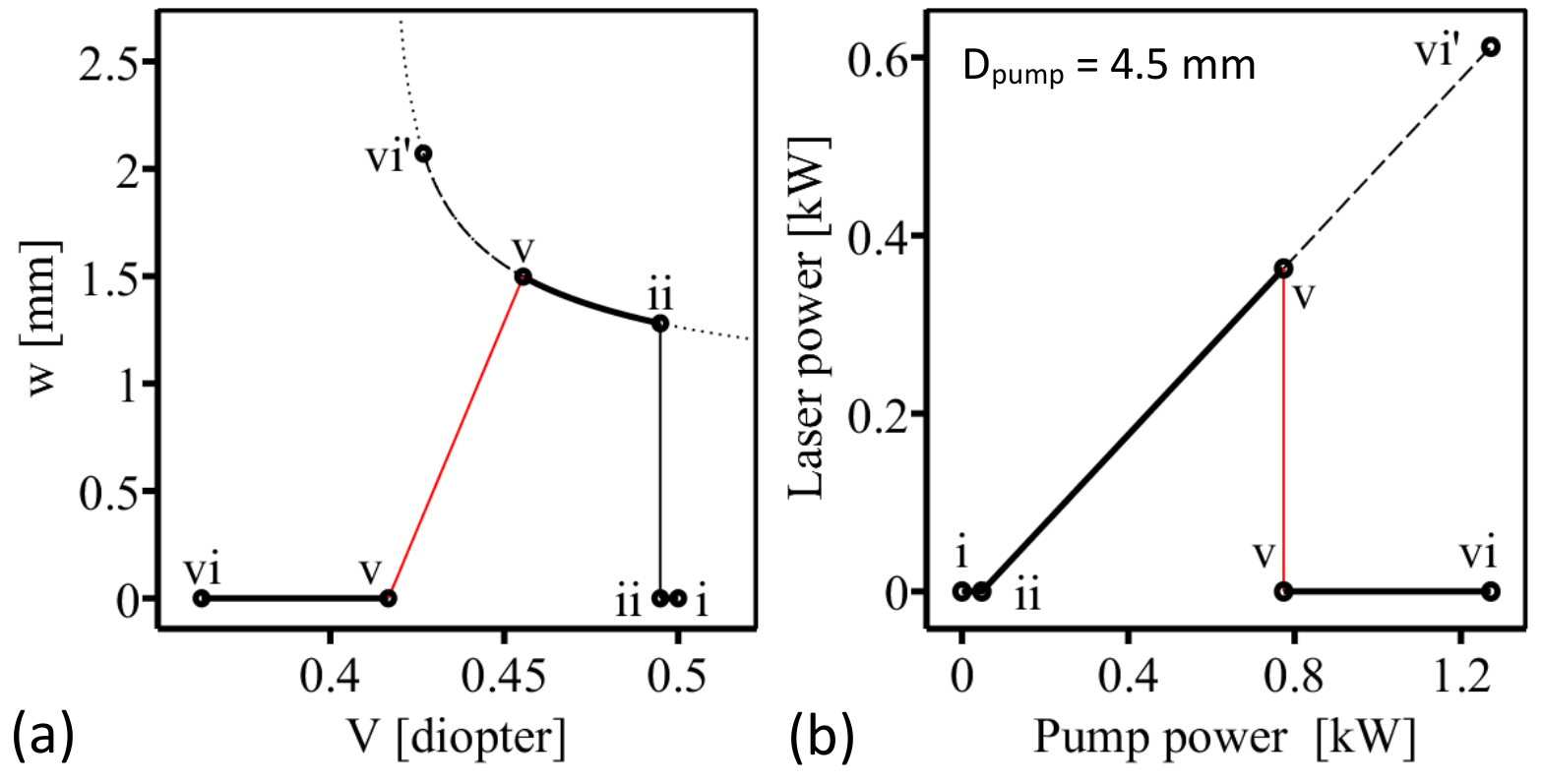}
  \caption{\label{fig:stability_layout2} (Color online) Similar to
    Fig.~\ref{fig:stability_layout1} but for a resonator with
    $L_1=500$~mm, $L_2=900$~mm, $L_3=100$~mm and
    $f_\mathrm{vex}=-1500$~mm. At position (v) the laser operation is
    disrupted by the onset of the thermal-induced misalignment
    mechanism and the dioptric power of the disk jumps from its laser
    operation value (0.45 diopters) to its fluorescence value (0.42
    diopters) as indicated by the red line. Therefore the maximal
    output power reachable is limited by the onset of the misalignment
    mechanism presented in this study.  The absence of laser operation
    between (i) and (ii) is because the gain at the disk does not overcome
    the losses at the out-coupler, while the absence of laser
    operation between (v) and (vi) is due to the thermal-induced
    misalignment. The points (vi') indicate waist and output
    power hypothetically reachable when neglecting the thermal-induced
    misalignment. }
\end{figure}

\begin{figure}[t!]
  \centering \includegraphics[width=0.85\linewidth]{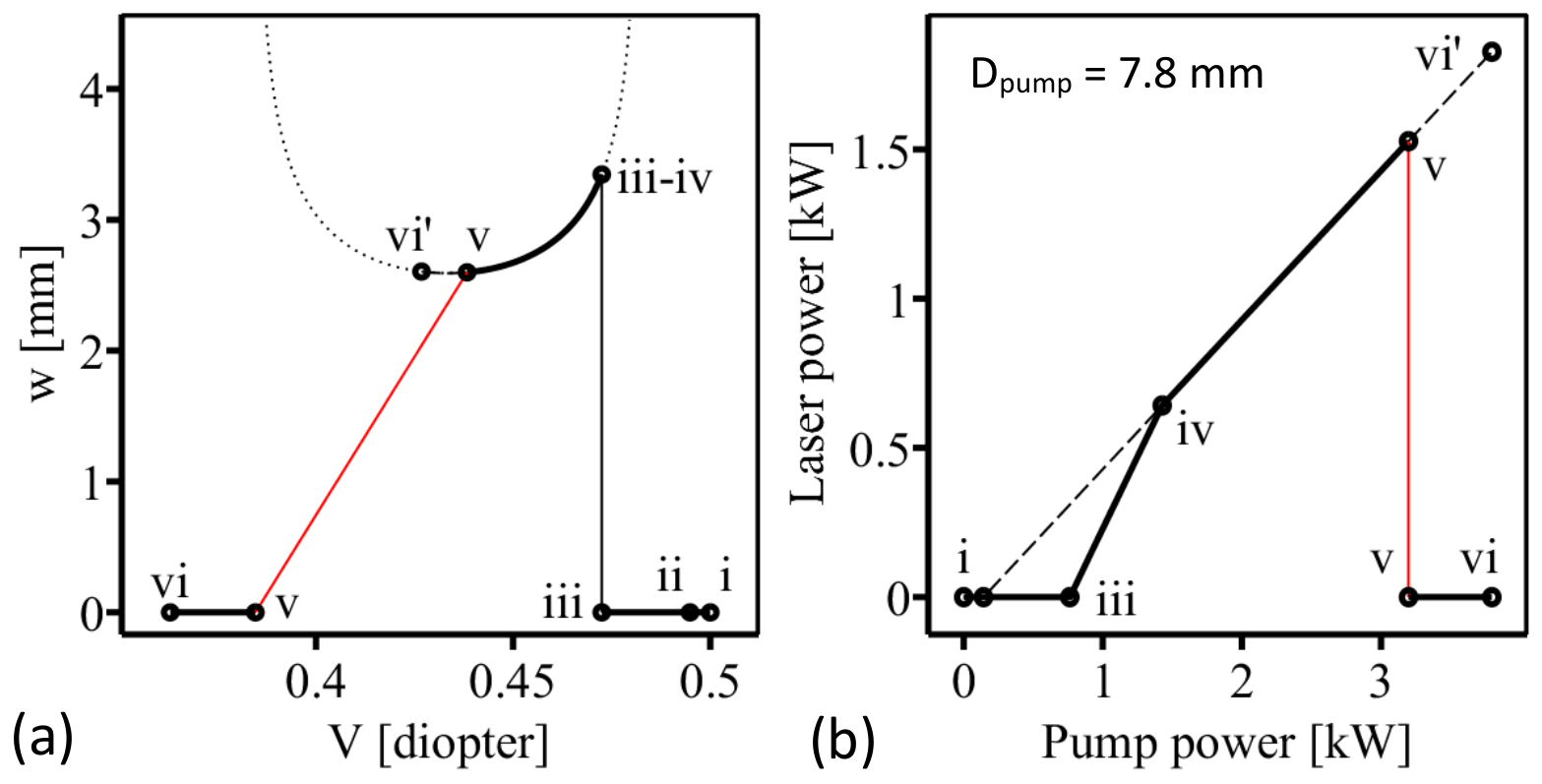}
\caption{\label{fig:stability_layout3} (Color online) Similar to
  Fig.~\ref{fig:stability_layout2} but for a resonator with
  $L_1=1000$~mm, $L_2=1600$~mm, $L_3=900$~mm and
  $f_\mathrm{vex}=-1000$~mm.  The absence of laser operation between
  (i) and (iii) has multiple origins: the gain at the disk does not
  overcome the losses at the out-coupler, or the resonator is outside
  the stability region, or the resonator is within the stability
  region but it has a waist $\mathrm{w} > 1.3\mathrm{w}_c$.  The
  absence of laser operation between (v) and (vi) is due to the
  thermal-induced misalignment. The behavior between (iv) and (v)
  assumes a slope efficiency of 50\% and a laser threshold of
  0.3~kW/cm$^2$. Between (iii) and (iv) there is a transition
  from fluorescence to laser operation.  }
\end{figure}

\begin{figure}[t!]
  \centering \includegraphics[width=0.85\linewidth]{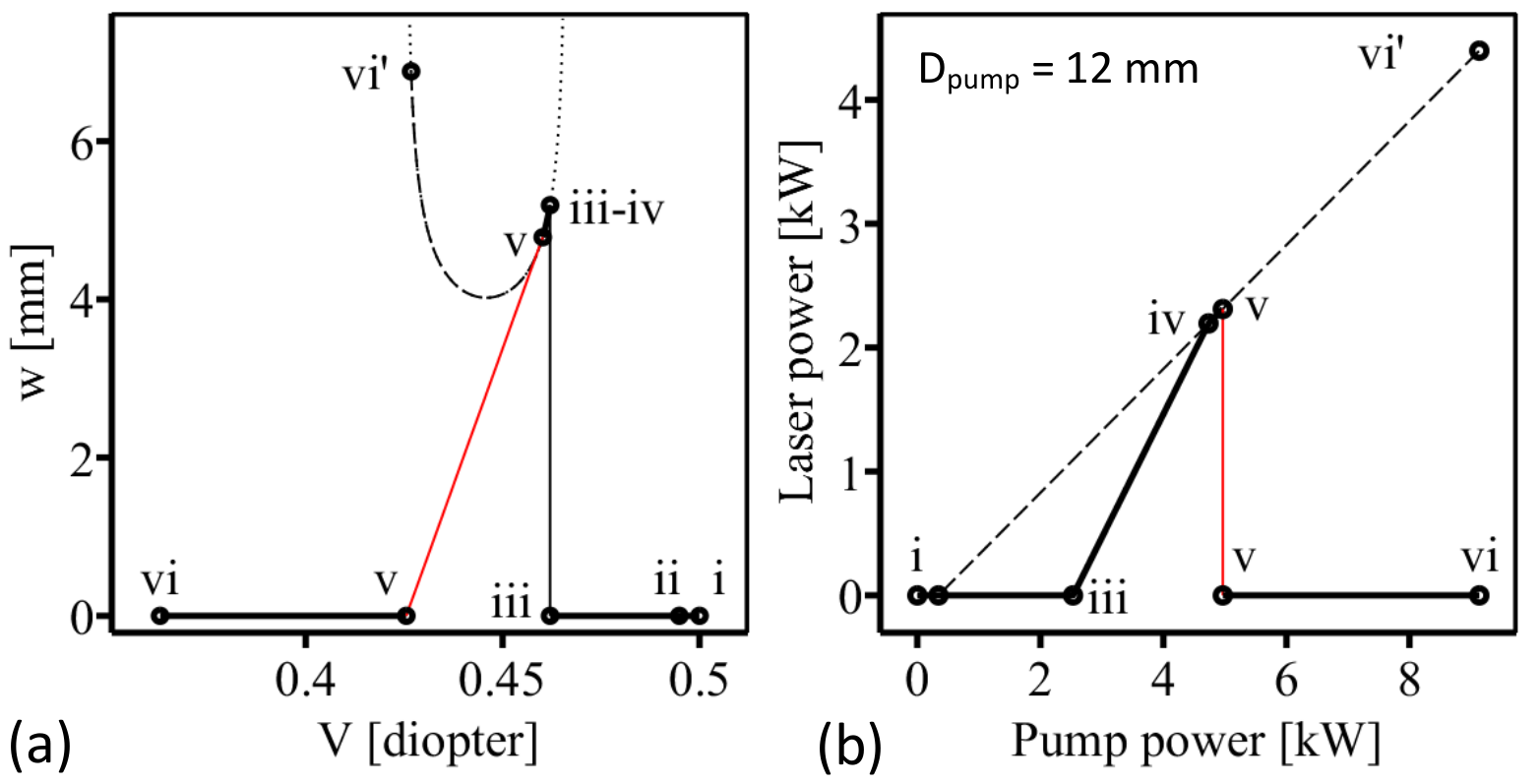}
\caption{\label{fig:stability_layout4} (Color online) Similar to
  Fig.~\ref{fig:stability_layout3} but for a resonator with $L_1=1000$~mm,
  $L_2=1600$~mm, $L_3=2000$~mm and $f_\mathrm{vex}=-750$~mm.  }
\end{figure}

For the modeling of the output power and the thermal lens effect
given in Figs.~\ref{fig:stability_layout1},
\ref{fig:stability_layout2}, \ref{fig:stability_layout3},
\ref{fig:stability_layout4} , \ref{fig:stability_layout5}
and \ref{fig:stability_layout6} we have assumed the  simplified
situation summarized in Table~\ref{tab1}.
\begin{table}[t!]
  \caption{\label{tab1} Parameters assumed to model the laser output
    power and the stability properties of
    Figs.~\ref{fig:stability_layout1}, \ref{fig:stability_layout2},
    \ref{fig:stability_layout3}, \ref{fig:stability_layout4},
    \ref{fig:stability_layout5} and \ref{fig:stability_layout6}. Notation: w$_c$
    is the waist of the eigen-mode in the center of the stability region, $P$ the pump power
    density and $V$ the dioptric power of the disk. }
\begin{center}  
{\renewcommand{\arraystretch}{1.3}
  \begin{tabular}{l l}
  \hline
  \hline
   Maximal pump power density       & 8 kW/cm$^2$    \\
   Pump power density at& \\[-1.7mm]
   laser threshold & 0.3 kW/cm$^2$   \\
   Slope efficiency                 & 50\%             \\
   Pump spot diameter       &$D_\mathrm{pump}=3\, $w$_c$  \\
   Disk dioptric power \\[-1.7mm]
   (unpumped) & $V(\mathrm{unpumped})=0.5$ 1/m\\
   Thermal dioptric power \\[-1.7mm]
   in fluorescence operation & $\frac{dV}{dP}=-0.017$ $\frac{\mathrm{1/m}}{\mathrm{kW/cm^2}}$~\cite{Perchermeier2013}\\
   Thermal dioptric power \\[-1.7mm]
   in laser operation & $\frac{dV}{dP}=-0.0092$ $\frac{\mathrm{1/m}}{\mathrm{kW/cm^2}}$ \cite{Perchermeier2013}\\
  \hline
  \hline
  \end{tabular}
}
\end{center}
\end{table}
The thermal dioptric power of disk in laser operation has been assumed
to be 50\% of the thermal dioptric power in fluorescence
operation~\cite{Chenais2006, Chenais2004, Perchermeier2013} with
$V_\mathrm{LB}>0$ (valid for TDL).
We further assumed that the dioptric power of the disk decreases
linearly with the pump power density, 
and that the laser operates in the fundamental mode.

\begin{table}[t!]
  \caption{\label{tab2} Description of significant states of the laser
    operation used in Figs.~\ref{fig:stability_layout1},
\ref{fig:stability_layout2}, \ref{fig:stability_layout3},
\ref{fig:stability_layout4}, \ref{fig:stability_layout5}
and \ref{fig:stability_layout6}.  }
\begin{center}  
{\renewcommand{\arraystretch}{1.2}
  \begin{tabular}{l l}
  \hline
  \hline
  i  & \parbox{0.85\linewidth}{
    The  disk is not pumped and its dioptric power  is $V=0.5$~1/m.}\\
  \hline\\[-4mm]
  ii & \parbox{0.85\linewidth}{The gain of the disk equals the losses at the out-coupler (other losses are neglected). Laser threshold is reached provided the resonator is within the ``classical'' stability region and the waist at the given dioptric power does not exceed 1.3 the layout value ( $\mathrm{w}\leq 1.3\, \mathrm{w}_c$).}\\[6mm]
  \hline\\[-4mm]
  iii & \parbox{0.85\linewidth}{The disk dioptric power  in {\it fluorescence} operation  gives rise to a stable resonator with $\mathrm{w} =1.3\, \mathrm{w}_c$. Laser operation is starting. }\\[2mm]
  \hline\\[-4mm]
  iv & \parbox{0.85\linewidth}{The  disk dioptric power in {\it laser} operation  is within the stability region  with $\mathrm{w} \approx 1.3\, \mathrm{w}_c$. }\\[2mm]
  \hline\\[-4mm]
  v & \parbox{0.85\linewidth}{The parameter $G$ becomes 1. Laser operation is disrupted giving rise to a rapid decrease of the disk dioptric power from the {\it laser} operation value to the {\it fluorescence} value (at the same pump power density).} \\[4mm]
  \hline
  vi & \parbox{0.85\linewidth}{Laser operation at the maximal allowed pump power density of 8~kW/cm$^2$. }\\
  \hline\\[-4mm]
  vi' & \parbox{0.85\linewidth}{Laser operation at the maximal allowed pump power density of 8~kW/cm$^2$ when the thermal induced misalignment mechanism  is ignored.  }\\[2mm]
  \hline
  \hline
  \end{tabular}
  }
\end{center}  
\end{table}

Figure~\ref{fig:stability_layout1} shows the stability plot and the
output power as a function  of the disk dioptric power  $V$ for
a resonator having an eigen-mode waist at the disk of about $1.5$~mm.
$V$ accounts for thermal and non-thermal (prior to
pumping) lens effects.
A qualitative understanding of the laser operation of this resonator
can be obtained by considering some particular states of the laser operation
indicated with Roman numbers from (i) to (vi)
as summarized  in Table~\ref{tab2}.

At zero pump power (i) the dioptric power of the disk is 
0.5~diopters (assumption).
With increasing pump power density the dioptric power decreases.
When the pump power density reaches 0.3~kW/cm$^2$ (ii) laser operation starts.
A further increase the pump power density leads to an increase of the
output power until the maximal allowed (due to optical damage) pump
power density of 8~kW/cm$^2$ (vi) is reached.
In this case, the laser resonator remains within the stability region
independently of the laser pump power density.
No limitations due to the above described thermal-induced misalignment are
noticeable in this layout.

In Fig.~\ref{fig:stability_layout2} the output power evolution and the
stability plot are given for a resonator layout whose stability region
is shifted compared with the layout of
Fig.~\ref{fig:stability_layout1}.
With increasing pump power and above the laser threshold (ii) the output
power increases.
However, at an output power of about 0.4~kW (v), laser operation stops because
the parameter $G$ becomes equal to 1.
As a consequence,   the thermal lens suddenly jumps
from the laser operation value to its fluorescence value (while the pump
power density remains constant).
A further increase of the pump power worsen the situation because $G$
further increases and the resonator moves out of the ``classical''
stability region.
Also in this case, provided that the laser operation would not be
disrupted by the thermal misalignment effect, the resonator would
stays inside the ``classical''~\cite{magni1986} stability region for
all pump power densities from 0 (i) to 8~kW/cm$^2$ (vi').
But for this layout the onset of the thermal-induced misalignment
limits the effective stability region and the maximal output power
as can be seen by comparing (v) with (vi').

Figure~\ref{fig:stability_layout3} shows the behavior of a resonator
having a larger eigen-mode waist of w$_c \approx 2.5$~mm (w$_c$
denotes the waist in the center of the stability region).
At point (ii) the disk gain becomes larger than the out-coupler losses, but
there is no laser operation because the resonator lies outside of the
stability region.
With increasing pump power, the resonator becomes stable
but laser operation starts only at point (iii) when the waist w of the
eigen-mode reaches a reasonable value that we assumed to be $\mathrm{w}=1.3\,\mathrm{w}_c$.
From (iv) to (v) the output power increases with the value given by the
assumed slope efficiency and laser threshold, while from (iii) to (iv)
a transition from fluorescence to efficient laser
operation occurs.
In this pump power density range the dioptric power of the disk can be
assumed to be constant as the heat load caused by the increase of pump
power density is compensated by the reduction of the heat load due to
the fast growing laser output power.

Also in this case the output power of the laser is limited by the
onset of thermal-induced misalignment instabilities which occurs at position (v)
when $G$ becomes equal to 1.
Here the thermal lens of the disk rapidly mutates from the laser to
the fluorescence value.
In this case, the obtainable maximal laser output power does not
significantly deviate from the value which could be obtained at the
maximal pump power density of 8~kW/cm$^2$ (vi').
However, for a resonator with larger eigen-mode  (see
Fig.~\ref{fig:stability_layout4}) the limitations induced by the
thermal-induced misalignment become substantial.

Laser operation of the resonator given in  Fig.~\ref{fig:stability_layout4}
follows the same dynamics as in Fig.~\ref{fig:stability_layout3} but
shows an increased limitation arising from the thermal-induced
misalignment  given the shrinkage of the ``classical''
stability region.
In fact it has been demonstrated~\cite{magni1986} that the width of
the stability region scales with $1/\mathrm{w}_c^2$ where $\mathrm{w}_c$ represents the
eigen-mode waist at the position of the thermal lens (disk).
Hence, stable operation of  high-power laser becomes
increasingly challenging.

The  thermal-induced misalignment effect presented here further worsens
the situation because a constant (in absolute terms and independent of
w) range of the stability region becomes unusable reducing the
``effective'' stability region.
This unusable range starts from the weak focusing edge of the
``classical'' stability region and has a width proportional to the
pump power density.
This unusable range with $G\geq 1$ within the ``classical'' stability
region is related to the commonly applied rule of thumb that a laser
resonator should be designed to remain inside the stability region
also when the laser cavity is blocked.

No stable laser operation is possible and power scaling reaches its
limit when the width of the ``classical'' stability region -- with
increasing mode size -- shrinks to the width of the unusable range
(due to thermal-induced misalignment instabilities).
A decrease of the pump power density would decrease the width of the
unusable range allowing the use of larger eigen-mode and pump spots
but the maximal output power would remain approximatively the same.

Possible non-linear variations of the thermal lens versus pump power
densities~\cite{Smrz} further amplifies the limiting effect of the here
disclosed effect.
However, proper resonator designs as exposed in the next section can
be used to circumvent this problem.

\begin{figure}[t!]
  \centering \includegraphics[width=0.65\linewidth]{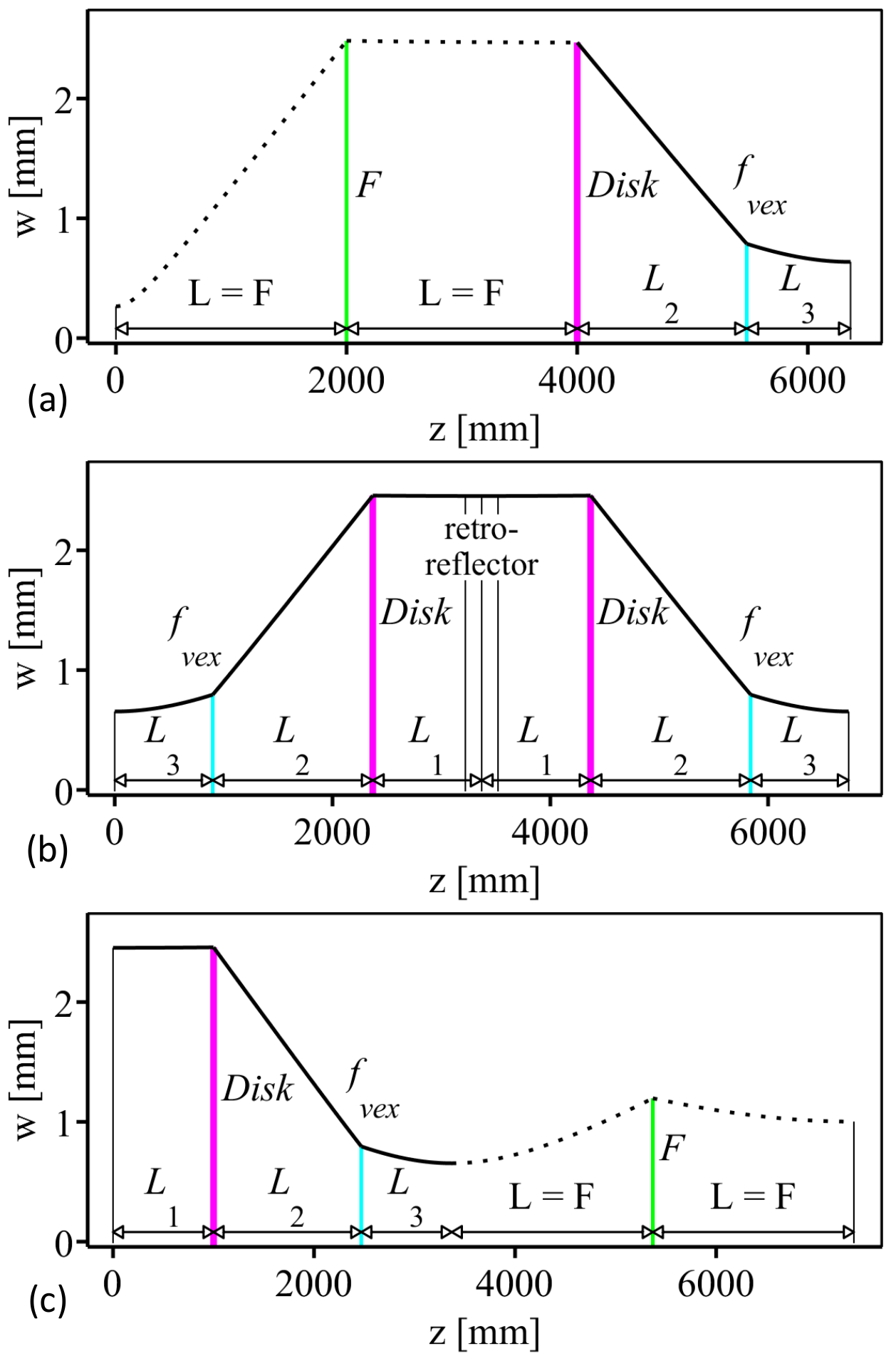}
\caption{\label{layout_2f} (Color online) Resonator architectures with
  $G<1$ (assuming $V_\mathrm{LB}>0$), i.e., unaffected by the
  thermal-induced misalignment. The vertical lines represent the
  position of the various optical elements. The eigen-mode waist w
  evolution along the resonator is also indicated. (a) This resonator
  has been obtained by replacing the left side ($L_1$) of the
  V-shaped resonator design of Fig.~\ref{fig:layouts} with a
  Fourier-transform segment. This is achieved by inserting a focusing
  element with focal length $F$ at a distance $F$ from the disk and
  from the end-mirror. (b) Resonator formed by combining two identical
  or quasi-identical optical segments, each of which having a layout
  of a V-shaped resonator as in Fig.~\ref{fig:layouts}. A
  retro-reflector is placed between the two segments and the same disk
  has to be used in both segments~\cite{Schuhmann2016}. (c) Similar to
  (a) but in this case the Fourier-transform segment is used to extend
  the right branch of the V-shaped resonator.  }
\end{figure}

\section{Resonator designs insensitive to the thermal-induced misalignment}
\label{sec:solution}
In this section we present three resonator architectures which avoid
the thermal-induced misalignment by keeping $G<1$ (assuming $V_\mathrm{LB}>0$).
The resonator shown in Fig.~\ref{layout_2f} (a) has been obtained by
replacing the  free propagation of length $L_1$ on the left side of
 Fig.~\ref{fig:layouts} by an optical
segment acting as a Fourier transform.
Since the ABCD-matrix of a Fourier transform based on a lens of focal
length $F$ reads
\begin{equation} 
  \left[
  \begin{array}{ c c }
     A_L & B_L \\
     C_L & D_L
  \end{array} \right]
=
  \left[
  \begin{array}{ c c }
     0 & F \\
     -1/F & 0
  \end{array} \right],
\end{equation}
the $G$ parameter of Eq.~(\ref{eq:G2}) becomes zero for
all layouts.
The physical origin of this stabilization arises from the fact that the
back and forth propagation of the beam in the Fourier segment
corresponds to a $4F$-relay imaging from pass to pass on the disk.
An on-axis laser beam reflected on the disk tilted by $\alpha_r/2$ leaves
the disk towards the Fourier segment with an angle $\theta_r=\alpha_r$.
When the beam is coming back to the tilted disk after the propagation
in the $4F$-relay imaging system its angle is inverted so that
$\theta_r=-\alpha_r$.
The subsequent reflection of the beam on the disk brings the beam back
on axis ($\theta_r=0$).
The beam thus reproduces itself on the right side of the resonator
independently of the disk tilt.

The drawback of this architecture is that for high power resonators
the cavity becomes exceedingly long.
This is caused by the requirement to have large beam waist at all
optical elements which calls for large $F$.

A retro-reflector (corner cube) as well inverts the beam angles: the
beam leaving the disk with angle $\theta_r=\alpha_r$ returns after the
reflection at the corner cube with an angle $\theta_r=-\alpha_r$.
Therefore also the layout of Fig.~\ref{layout_2f}~(b) is
stable against the thermal-induced misalignment.
Differently from the layout based on the Fourier extension, the
propagation from disk to disk via the corner cube can be short even
for large beam waists.

However, the beam offset generated by the corner cube prevents the
usage of the corner cube as resonator end-mirror.
For this reason in Fig.~\ref{layout_2f}~(b) the corner cube is used as
folding mirror between two reflections on the same disk.
This resonator layout corresponds thus to a multi-pass resonator (4
reflections on the same disk per round-trip) exhibiting a larger gain
compared to previous layouts.
Note that the realization of this layout with two different disks would not
provide any cancellation of the thermal-induced effect.
Moreover it has been demonstrated that the stability region of such
multi-pass resonators having only one disk does not depends on the
number of reflections at the disk and have been thus proposed to solve
present energy scaling of mode-locked laser
oscillators~\cite{Schuhmann2016} .

\begin{figure}[t!]
  \centering \includegraphics[width=0.85\linewidth]{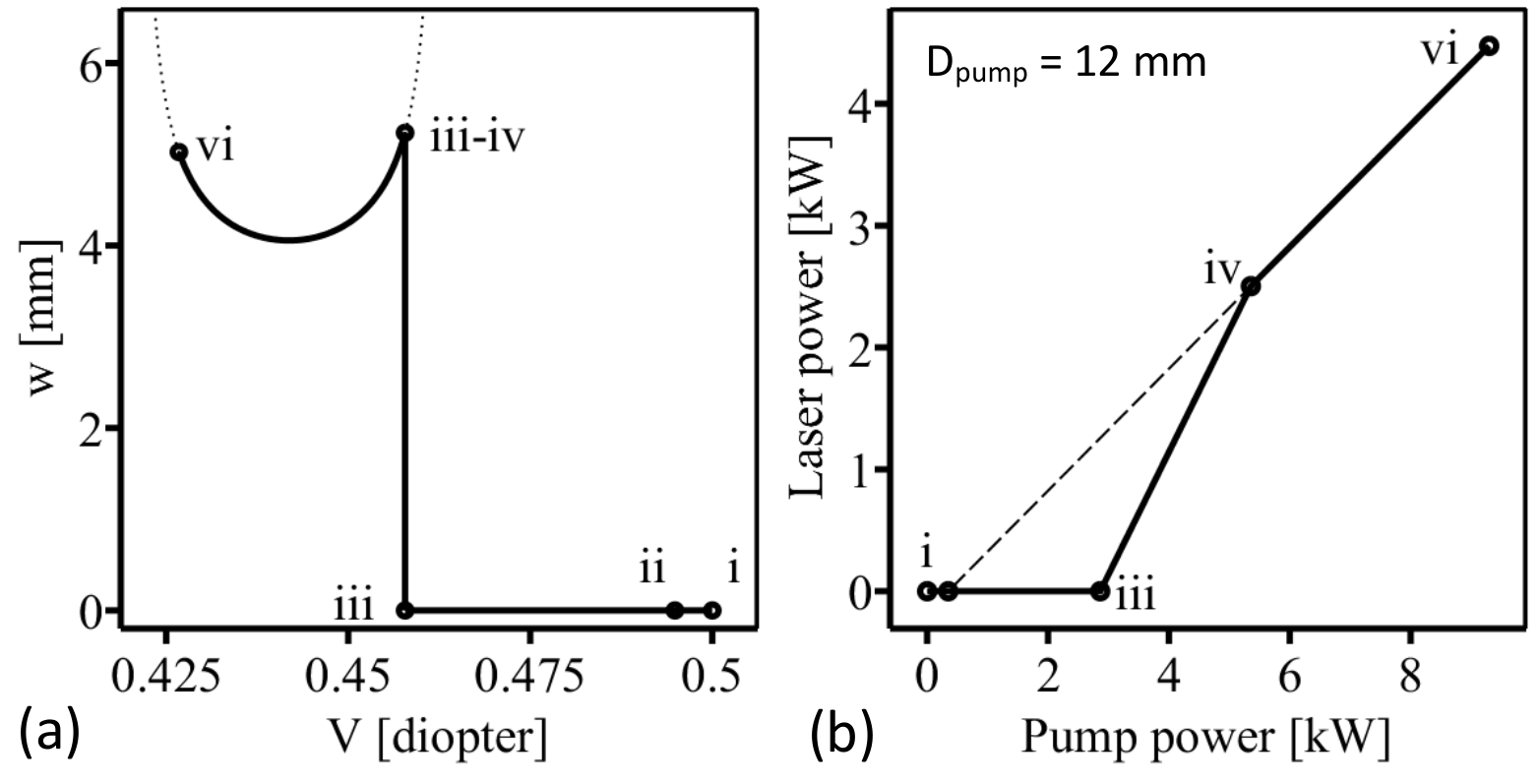}
\caption{\label{fig:stability_layout5} Similar to
  Fig.~\ref{fig:stability_layout4} but with a resonator design based
  on Fig.~\ref{layout_2f} with $L_2=1620$~mm, $L_3=2000$~mm and
  $f_\mathrm{vex}=-750$~mm. As for this resonator $G<1$ (assuming $V_\mathrm{LB}>0$), the
  thermal-induced misalignment presented in this study does
  not limit the maximal output power.  }
\end{figure}
\begin{figure}[t!]
  \centering \includegraphics[width=0.85\linewidth]{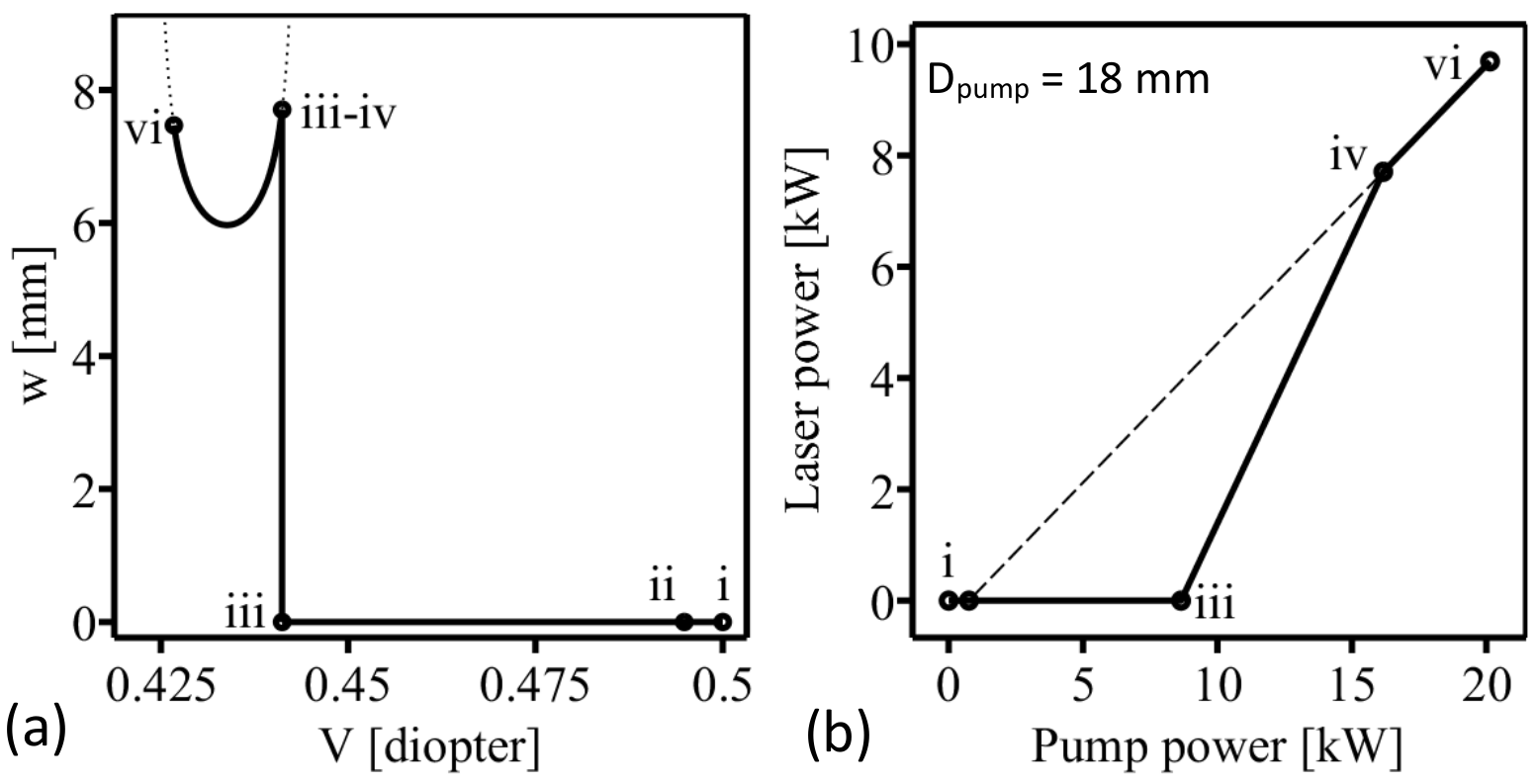}
\caption{\label{fig:stability_layout6}  Similar to
  Fig.~\ref{fig:stability_layout5} but with a resonator design 
   with larger eigen-mode waist ($L_2=1605$~mm, $L_3=5000$~mm and
  $f_\mathrm{vex}=-750$~mm). }
\end{figure}

In Fig.~\ref{layout_2f} (c) the Fourier-transform segment has been
added to the right side of the resonator of Fig.~\ref{fig:layouts}.
This addition leads to a sign change of $G$. In such a way $G$ becomes
negative stabilizing the resonator~\footnote{For active media having
  negative $V_\mathrm{LB}$, like rod lasers, this design leads to $G>0$
  limiting the power scaling. On the contrary the design given in
  Fig.~\ref{fig:layouts} having $G<0$ are power scalable for negative $V_\mathrm{LB}$.}.

The resonator adaptations presented in this section leads to larger
output power and larger ``effective'' stability ranges as can be seen
by comparing Fig.~\ref{fig:stability_layout4} with
Fig.~\ref{fig:stability_layout5}.
This opens the way for a further increase of the waist $\mathrm{w}_c$ resulting
in power scaling as shown in Fig.~\ref{fig:stability_layout6}.

\section{Conclusion}

We have exposed for the first time a fundamental obstacle to power scaling of TDL
related with self-driven growth of misalignment due to thermal-lens
effects.
We have found a parameter $G$ which serves to evaluate the response of
an optical resonator to an excursion of the laser eigen-mode at the
active medium position accounting for the changes of the OPD at the
active medium caused by the excursion of the laser eigen-mode itself.
This parameter $G$ can be computed using the ABCD-matrix formalism and
the knowledge of the active medium thermal lens in laser and
fluorescence operation.

When designing resonator layouts the region where $G\geq 1$ has to be
avoided.
This results for standard TDL design (V-shaped resonator) in a
restriction of the ``classical'' stability region (where stable laser
operation can be achieved).
Hence, it becomes particularly limiting for high-power TDL.

This limitation can be attenuated by reducing the thermal lens
difference $V_\mathrm{LB}$ between laser operation and fluorescence
operation which can be achieved by increasing the thermal conductivity
of the disk--heat-sink assembly, by increasing the stiffness of the
heat-sink and by lowering the heat deposition for example by using
zero-phonon line pumping~\cite{Smrz}.

However, this effect can be completely avoided by suited resonator
layouts as presented in Sec.~\ref{sec:solution} or by implementing an
active feedback as the beam excursion grows on a time scale of
milliseconds.

\section{Appendix: Simulations based on finite elements methods}

The thermal-induced OPD difference (at the disk) between fluorescence
(only pumped) and laser operation is at the core of this study.
The model we have developed describing this interplay and providing a
simple criterion to characterize the sensitivity of resonators to this
phenomenon assumes that a change of the eigen-mode position at the
disk generates a change of its OPD which corresponds
to a tilt of the disk.
The FEM simulations presented in Fig.~\ref{fig:OPD_pump_beam}
demonstrate the validity of this assumption for small beam excursion ($X\lesssim 1$~mm).
In this Appendix we specify the geometry and parameters entering  the FEM
simulations underlying the OPD profiles of Fig.~\ref{fig:pump} and
Fig.~\ref{fig:OPD_pump_beam}.

To account for the asymmetry caused by the off-axis laser beam 3D
simulations must be performed.
This differ from typical FEM simulations~\cite{Speiser2007,
  Speiser2009, Zhu2014} of TDL which usually assume rotational
symmetry and thus are performed only along a radial cut of the disk.
To reduce computing time however only half of the disk is simulated
and the appropriate symmetry conditions is used to extend the
simulations to the whole disk.
“Autodesk Simulation Mechanical 2015”  has been used.

The FEM simulations assume a diamond heat-sink of 1.5~mm thickness
and 25 mm diameter, and an Yb:YAG active material with 140~$\mu$m
thickness and a diameter of 20~mm.
The heat-sink is supported at its edge, while its backside is held at a
constant temperature of $T=13$~$^\circ$C.
Disk coatings and contacting layers have been neglected.

A flat-top pump beam of 12~mm diameter generates a heat load of
50~W/mm$^3$ in the active material  while it is assumed that in the
laser beam area (diameter of 9.6~mm) the laser beam reduces the heat
load by a factor of 2.
Such a reduction which strongly depends on running conditions, active
medium material and pump wavelength has been observed for example
in~\cite{Chenais2006, Chenais2004, Perchermeier2013, Smrz, Peters2009}.

\begin{figure*}[t!]
  \centering \includegraphics[width=1\textwidth]{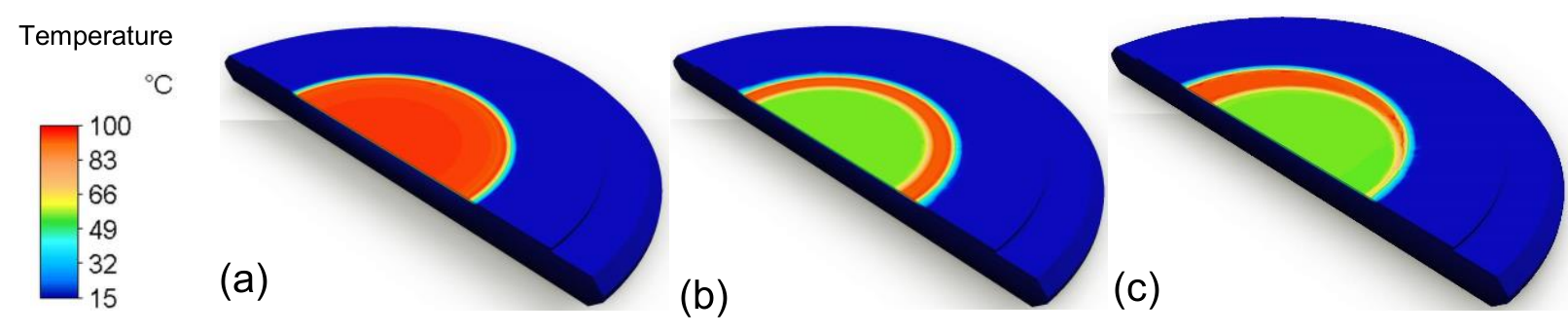}
  \caption{\label{fig:FEM} (Color online) Temperature distributions of
    the disk--heat-sink assembly with the geometry and heat loads as
    specified in the main text. (a) Temperature distribution caused by
    the pump only. The laser is operated in fluorescence mode. (b)
    Laser operation reduces the heat deposited at the resonator
    eigen-mode position. Here the laser eigen-mode is centered
    relative to the pumped area while in (c)  the laser
    eigen-mode is 1~mm off from the disk--pumped-area axis.  }
\end{figure*}
\begin{table}[t!]
  \caption{\label{tab3} Parameters assumed in the FEM simulations to
    model the thermal lens of the disk contacted to the diamond substrate. }
{\renewcommand{\arraystretch}{1.3}
  \begin{center}
  \begin{tabular}{l l}
  \hline
  \hline
  Yb:YAG (7\%) thermal conductivity     &  7  W/mK    \\
  Yb:YAG Young's modulus                &  300  GPa    \\
  Yb:YAG avg. thermal expansion         &  $8\times 10^{-6}$ 1/K   \\
  Yb:YAG refraction index change\\[-1.7mm]
  versus temperature ($dn/dT$)            & $9\times 10^{-6}$ 1/K    \\
  Diamond  thermal conductivity           &  1900  W/mK    \\
  Diamond  Young's modulus                &  1100  GPa    \\
  Diamond  avg. thermal expansion         &  $9\times 10^{-7}$ 1/K   \\
  \hline
  \hline
  \end{tabular}
  \end{center}
  }
\end{table}
Figure~\ref{fig:FEM} shows the temperature distribution at the disk
surface for three different conditions computed with FEM using the
parameters summarized in Table~\ref{tab3}.
In (a) the laser is in fluorescence mode, i.e., the active medium is
pumped but there is no laser light produced.
In transverse direction the temperature is constant within the pumped
area.
The heat-sink temperature is much smaller compared to the disk due to
the superior conductivity of the diamond relative to Yb:YAG.
In (b) the laser is operating in optimal conditions. The laser
eigen-mode is perfectly aligned with the disk--pumped-area axis.
As the circulating laser intensity reduces the thermal
load~\cite{Chenais2006, Chenais2004, Perchermeier2013, Smrz,
  Peters2009} the region of superposition between pumped-area and
laser eigen-mode is colder.
In (c) there is a 1~mm deviation between the laser eigen-mode axis and
the pumped region axis which leads to an asymmetric temperature
profile.
It is this asymmetric temperature distribution which causes an
asymmetric mechanical deformation of the disk backside which produces
the tilt effects described above.
Adding the axial expansion of the active medium and correcting for the
temperature dependence of the active medium refractive index we obtain
OPD profiles whose radial cuts are shown in Fig.~\ref{fig:pump} and
Fig.~\ref{fig:OPD_pump_beam}.

\section*{Funding Information}

We acknowledge the support from the Swiss National Science Foundation:
Projects SNF 200020\_159755, SNF 200021L\_138175, SNF 200021\_165854,
and the ERC StG. \#279765.


\end{document}